# Electro-Orientation and Electro-Rotation of Metallodielectric Janus Particles

Daniel Sofer[1], Gilad Yossifon[1,2*], Touvia Miloh[2]

[1]*Faculty of Mechanical Engineering, Micro- and Nanofluidics Laboratory, Technion – Israel Institute of Technology, Haifa 32000, Israel*

[2]*School of Mechanical Engineering, University of Tel-Aviv, Tel-Aviv 69978, Israel*

* corresponding author: gyossifon@tauex.tau.ac.il

**Abstract**

The electro-rotation (EROT) and electro-orientation (EOR) behavior of metallodielectric (MD) spherical Janus particles (JP) are studied analytically and verified experimentally. This stands in contrast to previous either heuristic or numerically computed models of JP dipoles. First, we obtain frequency-dependent analytic expressions for the corresponding dipole terms for a JP composed of a dielectric and metallic hemispheres, by applying the 'standard' (weak-field) electrokinetic model and using a Fourier-Legendre collocation method for solving two sets of linear equations. EROT and EOR spectra, describing the variation of the JP's angular velocity on the forcing frequency of a rotating and non-rotating spatially uniform electric field, respectively, are explicitly determined and compared against experiments (different JP size and solution conductivity). A favorably good qualitative agreement between theory and experimental measurements was found.

1. **Introduction**

Research on colloidal particles behavior has a long tradition. By determining their shape and physical properties, one can have a variety of mechanisms to manipulate a particle to execute a unique and controlled behavior. During the past several decades, the idea of breaking the symmetries (geometrical and/or physical) of a particle has played an essential role in micro-swimmer's research [1]–[5]. Breaking the symmetry clearly creates a preferable particle propulsion direction. One of the most straightforward ideas of symmetry- breaking is using a two-sided microsphere assembled from metallic and dielectric hemispheres, termed also as Janus particle (JP). A common technique to exploit this asymmetry is to apply an AC voltage to induce frequency-dependent multipoles within the JP and causing an asymmetric potential gradient near the particle. The electrical response of a JP can be estimated in terms of the corresponding leading-order multipoles. Besides the electrical force that may be applied on the induced dipole itself, the potential gradient forces the ions in the electrolyte to move and generate an asymmetrical electro-hydrodynamic force on the particle-electrolyte interface to electrokinetically drive the particle.

By enforcing an AC voltage, a cloud of ions accumulates near the interface of the JP polarizable hemisphere. It creates a layer of condensed ions, which is called a Debye layer or electric-double-layer (EDL). This layer manifests itself as a capacitor and depends on the AC frequency of the applying electric field. Applying different frequencies and using different electrode structures for controlling the externally applied electric field, one can employ different electrokinetic mechanisms for manipulating the particles, e.g., electro-hydrodynamics (EHD) [1], induced-charge electro-osmosis (ICEO) [1, 2], self-dielectrophoresis (sDEP)[5], dielectrophoresis (DEP)[6], [7], electro-rotation (EROT)[8]–[10] and electro-orientation (EOR) [4], [11]–[13]. Usually, a low-frequency electric field gives rise to EHD effects. The applied frequency is generally smaller than the relaxation time and thus provides enough time for the induced EDL and related ICEO to build up. In comparison, the high-frequency motion is usually a result of Maxwell-Wagner (MW) forces and torques that act on the induced dipole within the particle.

One of the main difficulties with analyzing JPs motion is to find the corresponding induced dipole behavior in terms of the inherent material JP asymmetry and forcing frequency. In spite of the fact that JPs have been heavily experimentally investigated, only few studies have focused on the theoretical investigation of their mobility behavior [8], [14], [15]. The built-in complexity of JPs has also led researchers to over-simplify the polarization mechanism model. For example, some works [2], [14] proposed to approximate JP polarization by the average of the corresponding polarization values of the two homogenous hemispheres (metallic and dielectric). This simplification led to the paradoxical conclusion that JPs do not have a preferable induced dipole direction, and hence while it may serve as a useful approximate for EROT response, it cannot be applied for example for EOR. Another work [15], describes the polarization of a JP as equivalent to the polarization of two un-related material hemispheres. As far as we know, only one recent attempt [8] has taken an analytical approach to solving the full problem and even then, was forced to solve it numerically by using a computer simulation. Such a numerical approach, aiming for example to determine the induced dipole in the direction parallel to the interface (y-direction in Fig.1), has its own drawbacks in terms of accuracy and computational cost. Boymelgreen and Miloh [22] computed analytically the dipoles of a metallodielectric JP, however, their solution was limited to the DC case and did not obtain the frequency dependent dipole term nor satisfied the charge conservation over the JP surface. Their approach was based on solving the interior electrostatic problem within the JP and matching it with the exterior solution using appropriate boundary condition both on the interior JP interface as well as on the JP surface. In contrast, our AC newly proposed method for MD-JP, based on using a Fourier -Legendre collocation (FLC) method, is much simpler and more direct in the sense that there is no need to solve the interior problem.

Another overlooked aspect of the current literature is the lack of explicit expressions for the real part of the induced dipole of JPs which governs its overall EOR response. Even though, the imaginary part of the JP dipole affecting its EROT dynamics has been studied before [8], [14], there are no previous studies, neither theoretical nor experimental about JP EOR mobility. Solving the EOR problem analytically, can be checked against actual JP electro-orientation measurements. A similar method has been also used before to determine the real part of the induced dipole of other particles [11]–[13], [16].

The theoretical methodology of the current problem is based on solving the linearized Poisson-Nernst-Planck (PNP) equations and applying the relevant boundary conditions imposed on the two hemispheres, using the 'standard' model based on assuming a 'weak-field', thin double layer and ignoring surface conductance effects (small Dukhin number). The above approach can be effectively used to explicitly determine both EROT and EOR spectra of a spherical MD -JP. EROT experimental data, can be achieved using a quadrupole electrodes structure with a phase shift of 90 degrees, as described in Fig. 1. This out-of-phase voltage generates a rotating electric field and induces an electric dipole within the JP. The phase lag between the induced dipole and the rotating electric field (due to the polarization mechanism of the JP) creates an electrostatic torque. The asynchronous EROT angular speed can then be found by equating the electrostatic torque to the viscous hydrodynamic (Stokes) torque (ignoring inertia effects), since ICEO effects on the angular velocity can be neglected with respect to the electric torque at least for thin EDL [24]. To complete the formulation with regard to EOR, one needs also to derive the real parts of the induced dipoles associated with the ambient in-phase AC electric field components (Fig.1 (b)), thus enabling us to explicitly determine the transient rotation (orientation) of the JP. The overall goal of this work is to present a comprehensive analytical framework for evaluating both the EOR and EROT spectra and compare them against experimental measurements of spherical JP response under various AC electric forcing.

## 2. Experimental setup

Figure 1 shows the quadrupolar electrode array experimental setup for EOR and EROT characterization of JPs suspended within KCl solutions of varying conductivities. To increase repeatability, the solution was introduced within the electrode chamber for at least 10 min before taking measurements to allow all of the JPs to sink to the bottom surface. During all experiments, one or few JPs were placed within the center of the electrode array to prevent JP's electrical interaction (e.g., chain formation). Various AC field frequencies with a sinusoidal waveform were applied using a function generator (33250A, Agilent). In order to minimize the side effects (e.g., electrolysis, Joule heating) of the electric field, a sufficiently low-amplitude AC field ($5V_{pp}$) that still provides a sensible effect was applied. To generate the rotating field ($5V_{pp}$), a four channels arbitrary waveform generator (TGA12104, AIM & Thurlby Thandar Instruments) with 90° phase-shift between the channels was connected to the four planar electrodes. To prevent from the heavier metallic side of the JP to face down due to gravity, we used two permanent magnets, symmetrically distanced (9cm) from the chamber so that the metallodielectric interface, wherein the metallic coating consists of a 10 nm thick Chrome followed by 50 nm thick ferro-magnetic Nickel (Ni) followed by 30nm/35nm thick gold (Au) for 10μm/5, 15μm JP diameter, respectively, was magnetically oriented parallel to the z-direction (where the quadrupolar electrode array is in the x-y plane) as described in Fig.1. The symmetrical configuration of the two magnets relative to the chamber and the large distance at which these were located (~9cm), results in an approximately uniform magnetic field. As another indication to the negligible magnetic field gradients is the fact that the JP is not moving under the presence of the magnetic field alone, as would be the case if these were strong enough to translate the JP via magnetophoresis [17].

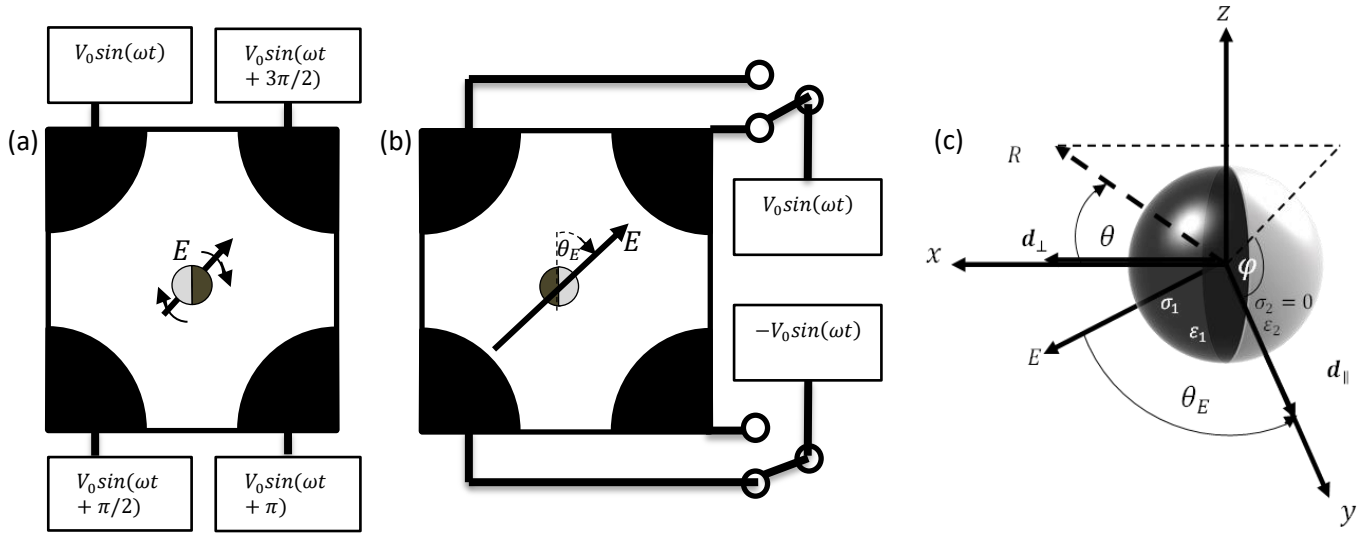

**Fig.1:** Experimental configuration of the quadrupolar electrode array for (a) electro-rotation (EROT) and (b) electro-orientation (EOR) experiments; $\theta_E$ depicts the angle between the electric field and the tangent to the metallodielectric interface direction. (c) Schematics of the JP geometry and definition of coordinate system as well as the induced dipoles and applied electric field.

The rotation of the JPs due to either electro-orientation or electro-rotation, was recorded using an Andor Neo sCMOS camera (20 frames per second) attached to a Nikon TI inverted epi-fluorescent microscope with a 20× (for 10 and 15 μm particles) and 40x (for 5μm particles) objectives lens and further analyzed by a particle tracking method using image processing software and self-developed image processing tools (ImageJ, NIH, OpenCV implemented in Python). Data shown in all results are the averaged values from at least 10 JPs (n ≥ **3**).

## 3. Theoretical modelling

We consider a spherical Janus consisting of two hemispheres of different complex permittivities $\tilde{\varepsilon}_1, \tilde{\varepsilon}_2$ embedded within an infinite symmetric monovalent electrolyte of complex permittivity $\tilde{\varepsilon}_e$. It is noteworthy that $\tilde{\varepsilon}_e = \varepsilon_e - i\frac{\sigma_e}{\omega}$ and $\tilde{\varepsilon}_j = \varepsilon_j - i\frac{\sigma_j}{\omega}$ ($j=1,2$) are referred to the electrolyte and JP's complex permittivities respectively, where $\varepsilon_e = 80\varepsilon_0$, $\varepsilon_1 = 6.9\varepsilon_0$ (for Au) [18], $\varepsilon_2 = 2.6\varepsilon_0$ (for Polystyrene)[19], with $\varepsilon_0$ representing the permittivity of a vacuum. The corresponding conductivity values are given by $\sigma_e = 0.2 \left[\frac{mS}{m}\right] (for\ DI)$, $0.6 \left[\frac{mS}{m}\right] (0.04mM)$, $15 \left[\frac{mS}{m}\right] (1mM)$ and $\sigma_1 = 4.26 \cdot 10^7$ (for Au)[20] and $\sigma_2 = 0$ (for Polystyrene) $\left[\frac{mS}{m}\right]$.

Let us next determine the corresponding dipole terms of a spherical MD JP due to uniform AC fields acting both along the x axis (normal to the JP interface), and along the y axis (parallel to the JP interface), as shown in Fig.1(c).

3.1. Axisymmetric case

For a uniform field (satisfying Laplace's equation) acting along the *x* axis (axisymmetric case), the normalized (by the ambient electric field $E_0$; unit amplitude) electric potential $\phi_\perp$ outside a (thin) EDL, can be written as [12], [21]:

$$\phi_\perp(R,\mu) = -\left(R + \frac{a^3}{2R^2}\right)P_1(\mu) + \sum_{n=1}^{\infty}\frac{D_n P_n(\mu) \cdot a^{n+2}}{R^{n+1}}, \quad (1)$$

Where $(R,\theta,\varphi)$ are the spherical coordinates (see Fig.1) with $\mu = \cos\theta$, $a$ denotes the radius of the JP, $P_n(\mu)$ represent the Legendre polynomials of degree $n$ and $D_n$ are coefficients (multipoles) to be determined. Employing charge conservation considerations[9] on the polarizable JP surface, relates the charging of the EDL (proportional to the time derivative of the difference between the potentials across the EDL) to the current in the bulk (proportional to the normal derivate of the potential on the surface). For time-harmonic (AC) excitation, the time derivative can be simply replaced by $-i\cdot\Omega$, which renders (see & 2.1 in [9]) the following mixed (Robin)- type boundary conditions applied on the two hemispheres:

$$\frac{\partial \phi_\perp}{\partial R} = \frac{1}{a}\begin{cases} -i\Omega_1(\phi_\perp - \phi_{\perp,1}) \\ -i\Omega_2(\phi_\perp - \phi_{\perp,2}) \end{cases} \quad \begin{cases} 1 > \mu \geq 0 \\ 0 > \mu > -1 \end{cases} @ R = a, \tag{2}$$

where, $\phi_{\perp,1}$ denotes the 'floating' potential inside the JP at the metallic hemisphere, and $\phi_{\perp,2}$ is the corresponding potential in the dielectric hemisphere.

The non-dimensional frequencies ($\Omega_{1,2}$) in (2) can be derived by solving the linearized PNP equations [12], [21] ,under the assumptions of a 'weak-field', thin EDL $\lambda_0 \ll a$ and ignoring surface conductance (small Dukhin number), resulting in:

$$\Omega_j = \frac{(\omega a \lambda_0 / D)\cdot(\lambda_0/\lambda)}{1 + \frac{\tilde{\varepsilon}_e}{\tilde{\varepsilon}_j}\cdot\frac{a}{\lambda_0}\cdot\left(\frac{\lambda_0}{\lambda}\right)^3}, \quad j = 1,2, \tag{3}$$

where $\omega$ is the physical (forcing) AC electric frequency and $D$ denotes the diffusivity parameter (assumed constant $D = 2\cdot 10^{-9}\left[\frac{m^2}{s}\right]$) of the symmetric electrolyte. Here, $\lambda_0$ (real) represents the traditional Debye scale (EDL thickness) and $\lambda$ (complex) is defined by the following relation [21] $1/\lambda^2 = 1/\lambda_0^2 - i\omega/D$.

Looking on the two extreme cases, namely where $\tilde{\varepsilon}_i \to 0, \infty$, corresponding to ideally insulating ('dielectric') and polarizable ('metallic') materials, equation (3) renders $\Omega_j = 0$ and $\Omega_j = (\omega a \lambda_0/D)(\lambda_0/\lambda)$, respectively. When making the assumption of a thin EDL for the low frequency regime (below the MW limit), one gets $\lambda_0/\lambda \approx 1$ and $\Omega_i = (\omega a \lambda_0/D)$ representing the RC frequency. For simplicity, let us assume here the case of a maximum contrast (MD), whereby one of the JP hemispheres is ideally dielectric and the other metallic, implying that $\Omega_2 \to 0$.

The value of the floating potentials in (2), can then be easily determined by imposing the condition of zero total current over the particle surface $S$ (externally to the EDL), i.e,

$$\int_S \frac{\partial \phi_\perp}{\partial R}\bigg|_S dS = 0. \tag{4}$$

Previous studies [22] did not take this charge conservation condition into account, in attempting to evaluate the induced dipole terms within the JP. Herein, we assumed that the JP is initially uncharged. Since we are looking for time-averaged (over a period) quantities for moderate to high frequencies (~0.1 to $10^4$ non-dimensional RC frequency), the integration of the Coulombic term associated with constant surface charge (or equivalently constant zeta-potential) taken over the JP surface, is not expected to yield a 'steady' electric torque.

At any giving point on the dielectric surface denoted by $S_2$, the zero current condition, namely, $\left.\frac{\partial \phi_\perp}{\partial R}\right|_{S_2} = 0$, automatically holds since $\Omega_2 = 0$ (2). On the other hand, in view of the above condition and by virtue of (2) and (4), the floating potential prevailing on the metallic surface $S_1$ under AC forcing, is given by:

$$\phi_{\perp,1} = \frac{1}{S_1} \int_{S_1} \phi_\perp \bigg|_{S_1} dS_1 = \int_0^1 \phi_\perp(a,\mu) \, d\mu. \tag{5}$$

Inserting (1) in (5) leads to

$$\phi_{\perp,1} = a \left( -\frac{3}{4} + \sum_{n=1}^{\infty} D_n \gamma_{n,0} \right), \tag{6}$$

where the coefficients $\gamma_{m,n}$ are defined as

$$\gamma_{m,n} = \gamma_{n,m} = \int_0^1 P_n(\mu) \cdot P_m(\mu) \, d\mu. \tag{7}$$

Letting $\Omega_2 \to 0$ and inserting (1) and (6) into (2), yields the following two sets of equations for the multipole terms:

$$-\sum_{n=1}^{\infty}(n+1)D_n P_n(\mu) = \begin{cases} -i\Omega_1\left(-\left(\frac{3}{2}\right)P_1(\mu) + \sum_{n=1}^{\infty}\left(D_n P_n(\mu)\right) - \phi_{\perp,1}\right) \\ 0 \end{cases} \quad \begin{cases} 1 > \mu \geq 0 \\ 0 > \mu > -1 \end{cases} @ R = a$$

(8)

Eq. 8 is then solved by using a Fourier-Legendre collocation (FLC) method by multiplying it by $P_m(\mu)$, integrating over each hemisphere and summing the resulting two equations. The FLC procedure finally renders the following infinite (single) set of equations:

$$\left(2(m+1) - i\,\Omega_1(1-(2m+1)\,\gamma_{m,0}{}^2)\right)\frac{D_m}{2m+1} = i\,\Omega_1\left(\frac{3}{4}(\gamma_{m,0} - 2\,\gamma_{m,1}) + \sum_{\substack{n=1 \\ n \neq m}}^{\infty} D_n\left(\gamma_{n,m} - \gamma_{n,0}\,\gamma_{m,0}\right)\right). \tag{9}$$

The resulting linear system of equations (9), can be readily solved by means of truncation to render the frequency-dependent coefficients (multipoles) $D_n(\Omega_1)$. Consequently, we use in (9) $N \leq 150$ terms until convergence is obtained. Table 1 shows the typical convergence of the dipole-term $D_1$ versus $N$ for $f = 5KHz$ and $f = 5MHz$ (the corresponding spectra is displayed in the supporting materials). It can be thus concluded that for $N \geq 50$, the solution of (9) uniformly converges.

3.2. Transverse case

In a similar way to (1), for a uniform AC field acting along the y axis parallel to the JP interface (asymmetric case), the normalized electric potential $\phi_\square$ existing in the solute outside the EDL, is taken as:

$$\phi_\square(R,\mu,\varphi) = -\left(R + \frac{a^3}{2R^2}\right)P_1^1(\mu)\cos\varphi + \sum_{n=1}^{\infty}\frac{D_n^1 P_n^1(\mu)\cdot a^{n+2}}{R^{n+1}}\cos\varphi, \tag{10}$$

where $P_n^1(\mu)$ represents the first- order associated Legendre polynomials of degree n, and $D_n^1(\Omega_1)$ are the corresponding coefficients (multipoles) to be determined. Note that the same boundary conditions also hold for this (transverse) case and equations (2),(3),(4) and (5) are left unchanged. However, due to the dependency of $\phi_\square$ on $\cos\varphi$ in (10), one can readily show by virtue of (6) that $\phi_{\square,1} = 0$ and thus following (2) one gets ($\Omega_2 \to 0$):

$$\frac{\partial \phi_\square}{\partial R} = \begin{cases} -i\Omega_1\phi_\square/a \\ 0 \end{cases} \quad \begin{cases} 1 > \mu \geq 0 \\ 0 > \mu > -1 \end{cases} @\, R = a. \tag{11}$$

Using (11) and (10), and going again through the same FLC procedure, we obtain the following infinite set of linear equations:

$$\left(2(m+1) - i\,\Omega_1\right)\frac{m(m+1)D_m^1}{2m+1} = i\,\Omega_1\left(-\frac{3}{2}\gamma_{m,1}^1 + \sum_{\substack{n=1 \\ n \neq m}}^{\infty} D_n^1\,\gamma_{n,m}^1\right), \tag{12}$$

where the coefficients $\gamma_{m,n}^1$ are defined, in a similar manner to (7), by

$$\gamma_{m,n}^1 = \gamma_{n,m}^1 = \int_0^1 P_n^1(\mu) \cdot P_m^1(\mu) \, d\mu. \tag{13}$$

Convergence of the corresponding dipole term $D_1^1$ is again obtained for $N \leq 150$ as shown in Table 1 for the selected frequencies of $f = 5\,KHz$, and $f = 5\,MHz$, taking $a$ to be 5μm (radius). Determining the dipole terms in (9) and (12), depends on the higher-order multipoles. The full spectra for both the normal and transverse dipoles, are displayed in the supporting material. As far as we know, this is the first time that the corresponding higher-order frequency-dependent multipoles for a MD JP have been analytically computed.

| N | $D_1\ (@f = 5MHz)$ | $D_1^1\ (@\ f = 5MHz)$ | $D_1\ (@\ f = 5KHz)$ | $D_1^1\ (@\ f = 5KHz)$ |
|---|---|---|---|---|
| 2 | $1.512 - i0.012$ | $1.501 - i0.001$ | $0.170 - i0.214$ | $0.644 - i0.509$ |
| 5 | $0.953 - i0.084$ | $1.300 - i0.019$ | $0.167 - i0.177$ | $0.656 - i0.488$ |
| 10 | $0.459 - i0.124$ | $1.232 - i0.012$ | $0.171 - i0.172$ | $0.6579 - i0.4727$ |
| 50 | $0.399 - i0.022$ | $1.053 - i0.007$ | $0.173 - i0.169$ | $0.6594 - i0.469$ |
| 100 | $0.379 - i0.010$ | $1.036 - i0.004$ | $0.173 - i0.169$ | $0.6594 - i0.4687$ |
| 125 | $0.373 - i0.003$ | $1.036 - i0.009$ | $0.173 - i0.169$ | $0.6594 - i0.4687$ |
| 150 | $0.373 - i0.003$ | $1.031 - i0.003$ | $0.173 - i0.169$ | $0.6594 - i0.4687$ |

**Table.1:** Values of $D_1$ and $D_1^1$ at $f = 5KHz$ and $f = 5\,MHz$, as a function of N.

### 3.3 Electro-rotation (EROT) of JP:

Consider a JP subjected to a rotating electric field as described in Fig.1(a), which can be expressed in terms of its phasor as:

$$\mathbf{E}_{ROT}(t) = E_0 \Re\left[ (\mathbf{e_x} - i\mathbf{e_y})e^{-i\omega t} \right]. \tag{14}$$

where $(\mathbf{e_x}, \mathbf{e_y})$ denote unit vectors in the x and y directions respectively and $\Re$ stays for the real part. As demonstrated in previous studies [12], [21], [23], applying an AC ambient electric field exerts both electrostatic and electro-hydrodynamic torques on the JP. Those torques are usually referred to as electrostatic induced-dipole torque (IDT) and 'induced-charge-electro-phoretic' (ICEP) torques, respectively [24], [25]. In addition to these, a viscous (Stokes) torque is also applied by the liquid on the particle Interface.

The time-averaged (over a single frequency) IDT torque acting on the JP, can be accordingly expressed in terms of the induced dipole vector $\mathbf{p}$ as [26]:

$$\boldsymbol{\tau}_{IDT} = \tfrac{1}{2}\Re[\mathbf{p} \times \mathbf{E}^*] = 2\pi\varepsilon_e a^3 \Re\big[d_\perp E_x E_y{}^* - d_\| E_y E_x{}^*\big]\mathbf{e}_z. \tag{15}$$

where (*) denotes the conjugate and $\mathbf{p}$ can be written as:

$$\mathbf{p} = 4\pi\varepsilon_e a^3 [d_\perp E_x \mathbf{e}_x + d_\| E_y \mathbf{e}_y] \tag{16}$$

Here $d_\perp, d_\|$ represent the perpendicular and parallel induced-dipole complex polarizabilities, respectively, where following (1) and (10) one gets $d_\perp = -\tfrac{1}{2} + D_1; \quad d_\| = -\tfrac{1}{2} + D_1^1$.

Substituting (14) in (15) yields:

$$\boldsymbol{\tau}_{IDT} = -2\pi\varepsilon_e a^3 E_0{}^2 \Im[d_\perp + d_\|]\mathbf{e}_z. \tag{17}$$

where $\Im$ represents the imaginary part. For a steadily rotating sphere near a wall, the viscous torque can be estimated following [27] as:

$$\boldsymbol{\tau}_{\text{Viscous}} = -8\pi\eta a^3 \dot{\boldsymbol{\theta}} \left( \zeta(3) - 3\left(\frac{\pi^2}{6}-1\right)\left(\frac{d}{a}-1\right) + \ldots \right), \tag{18}$$

where $\dot{\boldsymbol{\theta}}$ is the angular velocity in the z-direction, $\zeta(3)$ is the Riemann zeta function, ($\zeta(3) = 1.20206$), $d$ is the distance of the JP's center from the wall, and $\eta$ is the dynamic viscosity of the solution (assuming $\eta = 1 \cdot 10^{-3} \, [Pa \cdot s]$).

For the case where the JP lies in close proximity to the wall (i.e., $d \approx a$), the viscous torque (18) can be written as:

$$\boldsymbol{\tau}_{\text{Viscous}} = -8\pi\eta a^3 \zeta(3)\dot{\boldsymbol{\theta}}. \tag{19}$$

The IDT angular velocity in this case is thus given by:

$$\dot{\boldsymbol{\theta}}_{IDT} = -\frac{\varepsilon_e E_0{}^2 \Im[d_\perp + d_\|]}{4\eta\zeta(3)}\mathbf{e}_z. \tag{20}$$

The minus sign, shows that the particle rotates in the opposite direction (counter-field) to the applied electric field.

Following [12], the general expression for the ICEP torque can be written as:

$$\boldsymbol{\tau}_{ICEP} = -\frac{3\varepsilon_e}{8a}\mathbf{e}_z \int_S Q\left(\bar{\mathbf{r}} \times \nabla\chi^*\right) ds. \tag{21}$$

Here, $\bar{\mathbf{r}} = (x, y)$ denotes the radius vector and the effective potential $\chi$ can be described as:

$$\chi = 2E_0 \left(\chi_1 - i \cdot \chi_2\right), \tag{22}$$

where, within the dipole-approximation, $\chi_1$ and $\chi_2$ are given by:

$$\chi_1 = \left(-1 + \frac{d_\perp a^3}{R^3}\right)x + \ldots \quad ; \quad \chi_2 = \left(-1 + \frac{d_\parallel a^3}{R^3}\right)y + \ldots . \quad (23)$$

The non-dimensional induced-charge density in (20), can be written as [21]:

$$Q = \left(\frac{\lambda}{\lambda_0}\right)^2 \left((\chi_1 - \chi_0) - i\chi_2\right) \quad @ R = a , \quad (24)$$

where,

$$\chi_0\big|_{R=a} = \frac{a}{2}(-1 + d_\perp). \quad (25)$$

Taking equations (21)-(25) all together, and using the dipole approximation in (23) gives:

$$\boldsymbol{\tau}_{ICEP} = \frac{3\pi\varepsilon_e a^3 E_0^2}{4\eta\zeta(3)}\left(\frac{\lambda}{\lambda_0}\right)^2 \Im\{(1-d_\perp)(1-d_\parallel^*)\}\boldsymbol{e_z}, \quad (26)$$

Which by using (18), renders the following expression for the sought ICEP angular velocity:

$$\dot{\boldsymbol{\theta}}_{ICEP} = \frac{3\varepsilon_e E_0^2}{32\eta\zeta(3)}\left(\frac{\lambda}{\lambda_0}\right)^2 \Im\{(1-d_\perp)(1-d_\parallel^*)\}\boldsymbol{e_z}. \quad (27)$$

Finally, using the super-position principle, we can accordingly determine the electric torque and ICEP combined contributions to the overall angular velocity of a rotating JP under EROT:

$$\dot{\boldsymbol{\theta}}_{EROT} = \left[-\frac{\varepsilon_e E_0^2 \Im[d_\perp + d_\parallel]}{4\eta\zeta(3)} + \frac{3\varepsilon_e E_0^2}{32\eta\zeta(3)}\left(\frac{\lambda}{\lambda_0}\right)^2 \Im\{(1-d_\perp)(1-d_\parallel^*)\}\right]\boldsymbol{e_z}. \quad (28)$$

Fig.3 shows the EROT spectra for IDT and ICEP separately as well as the combined value. ICEP contribution is modified here to 10% of its real value following the experimental results of [28]–[32], claiming that Stern layer effects, surface roughness and counter-ion crowding, all result in a considerable decrease in the measured ICEP velocity compared to the theoretical prediction.

### 3.4 Electro-orientation (EOR) of JP:

Next, we consider a JP under the same conditions (an infinite symmetric monovalent electrolyte), subjected to an arbitrary uniform electric field as described in Fig 1 (b. c). One can describe this EOR electric field by using its phasor as:

$$\mathbf{E}_{EOR} = E_0 \cdot \left[\sin(\theta_E)\mathbf{e_x} + \cos(\theta_E)\mathbf{e_y}\right]e^{-i\omega t}, \quad (29)$$

Where $\theta_E$ denotes the instantaneous angle between the applied electric field and the tangent direction to the JP interface (see Fig.1(b)). The time-averaged electrical torque exerted by the induced dipole (IDT) (within the above dipole approximation). is thus given by:

$$\boldsymbol{\tau}_{IDT} = -\pi\varepsilon_e a^3 E_0^2 \sin(2\theta_E)\Re(d_\parallel - d_\perp)\boldsymbol{e_z}. \quad (30)$$

The IDT angular velocity can then be written as:

$$\dot{\boldsymbol{\theta}}_{IDT} = -\Gamma_e \sin(2\theta_E)\boldsymbol{e_z}; \quad \Gamma_{IDT} \triangleq \frac{\varepsilon_e E_0^2 \Re(d_\parallel - d_\perp)}{8\eta\zeta(3)}. \tag{31}$$

In order to compute the ICEP torque in the case of EOR, one needs to use again (21)-(25), but instead of (22), we let following (29):

$$\chi = 2E_0\left(\sin(\theta_E)\chi_1 + \cos(\theta_E)\chi_2\right). \tag{32}$$

Replacing (22) with (32), we can determine the ICEP torque acting on a JP subjected to EOR forcing (29) as:

$$\boldsymbol{\tau}_{ICEP} = -\frac{3\pi\varepsilon_e a^3 E_0^2}{8\eta\zeta(3)}\left(\frac{\lambda}{\lambda_0}\right)^2 \sin(2\theta_E)\Re\{(1-d_\perp)(1-d_\parallel^*)\}\boldsymbol{e_z}. \tag{33}$$

Hence, the corresponding angular velocity due to ICEP is given by:

$$\dot{\boldsymbol{\theta}}_{ICEP} = -\Gamma_{ICEP}\sin(2\theta_E)\boldsymbol{e_z}; \Gamma_{ICEP} = \frac{3\varepsilon_e E_0^2}{64\eta\zeta(3)}\left(\frac{\lambda}{\lambda_0}\right)^2 \Re\{(1-d_\perp)(1-d_\parallel^*)\}. \tag{34}$$

Finally, it is possible to evaluate the combined IDT(31) and ICEP (34) contributions (dipole approximation) and express the overall EOR angular velocity of a JP, simply as:

$$\dot{\boldsymbol{\theta}}_E = -\Gamma\sin(2\theta_E)\boldsymbol{e_z}; \Gamma = \frac{\varepsilon_e E_0^2 \Re\{d_\parallel - d_\perp\}}{8\eta\zeta(3)} + \frac{3\varepsilon_e E_0^2}{64\eta\zeta(3)}\left(\frac{\lambda}{\lambda_0}\right)^2 \Re\{(1-d_\perp)(1-d_\parallel^*)\}. \tag{35}$$

The explicit solution for the ODE in equation (35) is:

$$\tan\theta_E(t) = \tan\theta_0 \exp(-2\Gamma t). \tag{36}$$

Here, $\theta_0$ is found by implementing the relevant initial condition of the system at $t=0$, namely $\theta_0 = \theta_E(0)$ and $\Gamma$ is defined in (35). Eq. (36) analytically determines the transient electro-orientation motion of a spherical MD JP under the combined effect of IDT and ICEP. Fig.4 shows the EOR spectra corresponding to IDT+ ICEP.

## 4. Results and Discussion
### 4.1. Tangential and normal induced electrical dipoles of a metallodielectric JP.

The induced dipole of a spherical JP consisting of two hemispheres (different material properties), subjected to arbitrary AC electric forcing, has never been computed analytically (see Fig 3). As mentioned before, the current literature suggests only some intuitive approximations, such as the averaged induced dipole of two different polarizable homogenous (metallic and dielectric) particles. This approximation may serve as a partial explanation for JP's motions and related spectra under EROT excitation. Nevertheless, since the induced dipole in these two cases (i.e., parallel and perpendicular to the JP interface) is the same, this approximation cannot predict the experimentally observed EOR behavior of a spherical MD JP, because the difference between the corresponding two dipoles is null! (31). Moreover, our new approach demonstrates that the RC frequency of the JP is expected to be higher than that for homogenous-metallic particles. This finding was also experimentally found and mentioned without explanation earlier by [10].

From Fig.2, we can see that the real part of the induced dipoles is negative at low frequencies and goes to positive values at higher values only in the tangential dipole case. The imaginary part of the induced dipoles vanishes both at low and high frequencies, as expected. Only in the intermediate frequency region, this value is non-zero (positive). This kind of a qualitative behavior, is typical for micro-particles [4], [11]–[13]. A common explanation for the particular shape of this curve, is related to the fact that ions tend to screen the induced dipole of the JP at low frequencies. However, the ions do not have sufficient time to respond to the applied field at high frequencies. Therefore, the theoretical prediction of the induced dipole is considerably larger, affecting the electrostatic torque exerted on the JP particle.

It is also worth noting, that both the real and the imaginary parts of the parallel induced dipole are higher than the perpendicular induced dipole. It is thus plausible to assume that the charges of each hemisphere are 'locked' in their hemisphere and can't skip to the other hemisphere. So, each hemisphere has both a long direction and a short direction. Thus, the dipole, which is proportional to the distance between the charges, is higher in direction tangent to the MD interface.

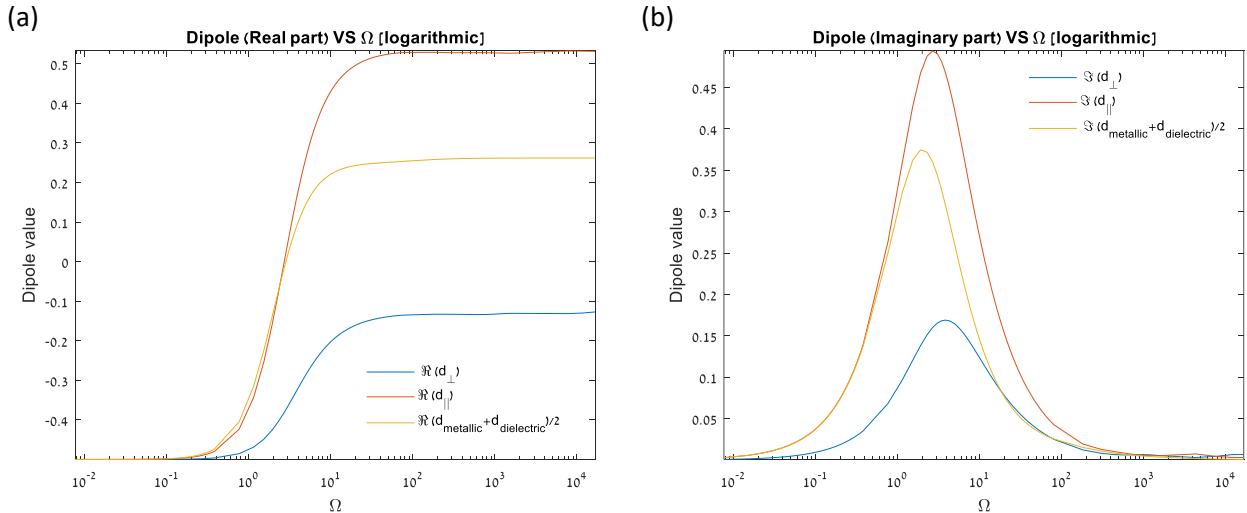

**Fig.2:** Theoretically calculated real (a) and imaginary (b) parts of the electrically induced dipoles acting in the directions tangential and normal to the MD JP interface versus the applied frequency.

## 4.2. Electro-rotation of a JP

By computing the values of the induced-dipoles for the transverse and axisymmetric cases, using equations (19) and (26), the theoretical IDT and ICEP angular velocities can be explicitly obtained. In a similar manner to [11], we used for our experimental setup the same analytical-numerical computed ratio value of the electric field magnitude, as found from their corresponding COMSOL simulations, in order to obtain a better accuracy. Therefore, instead of simply taking $E_0 = \frac{\Delta V}{\Delta d}$, we use instead $E_0 = c_i \frac{\Delta V}{\Delta d}$, where $E_0$ is the discussed electric field magnitude, $\Delta V = 5\,[V]$ is the corresponding voltage drop set to the electrodes, $\Delta d = 500\,[\mu m]$ is the distance between the two diagonal electrodes and $c_i$ is the correction factor ($c_1$ for EROT $c_2$ and EOR). As a result, one gets $E_0^{(EROT)} = 7483\,[V/m]$, $E_0^{(EOR)} = 7778\,[V/m]$. The combined effect of both IDT and ICEP on EROT spectra is described in equation (27).

We normalize all results for the combined EROT spectra by $\varepsilon_e E_0^2 / \eta$ and present in Fig. 3, both analytical and experimental results. Note that we use a factor of 0.1 for the ICEP curve, suggested by [28] and attributed to Stern layer effects, which tends to reduce the actual slip velocity on the JP surface [28]. This correction factor associated with the voltage drop over the Stern layer, was first proposed by [30] to account for the observed discrepancy between experimental and theoretical ICEO predictions (see also [31][32]). The model is based on dividing the theoretical values by 1+δ, where δ denotes the ratio between the capacitance of the compact (Stern) to the diffuse (EDL) layers. Here, we have simply used the approximate value of 10 which has been proposed in several of these works, resulting in the above 0.1 correction factor, without attempting to use this correction factor as a fitting parameter.

Inspection of the data in Fig. 3, shows that the theoretical results can qualitatively explain the experimentally observed frequency dependence. Although qualitatively we get a reasonable agreement, yet we still do not have a plausible explanation for the quantitative mismatch of the maximum values between the experimental and analytical curves, however it is worth mentioning that similar differences were also reported in [14].

Another interesting finding, was the high correlation of the RC frequency, found between the measured data and the analytical curve. Notably, this correlation is visible only when using the normalized frequency. The new analytical methodology suggested here, gives a slightly better agreement (compared for example to [10]), between the analytical RC frequency $\Omega_{RC}^{Analytical} = 2.896$ and the approximate empirical value $\Omega_{RC}^{Empirical} = 3.05^{\pm 0.76}$.

Based on above analytical considerations, the amplitude of the EROT spectra does not depend on the JP's size (28), whereas experiments show a non-monotonic size-dependence (see Fig. 3). It is also worth mentioning that in a former investigation [10], it was found that the maximum velocity increases as the JP's size is reduced. On the other hand, other studies showed no monotonic relation between metallic [9], [29] and semiconducting [3] particle's size and EROT maximum rotation speed. Some works [3], [29] a non-linear dependence between the particle size and the maximum angular speed. A possible explanation for the dependency on particle size, may be connected to the difference in the surface charge [33] or to the surface conductance of different particles, which can vary substantially over the surface of non-metallic particles, e.g., semiconducting and JP. The particular size dependent problem of a MD spherical JP will be further analyzed in a future study.

As depicted in Fig.3(b), the results of the EROT spectra for different solution conductivities (DI water: $0.2\left[\frac{mS}{m}\right]$), 0.04mM KCl ($0.6\left[\frac{mS}{m}\right]$) and 1mM KCl ($15\left[\frac{mS}{m}\right]$), which support the well-known dependency of the RC frequency on the solution conductivity, are in agreement with previous studies [3], [8]–[12], [14], [23]. Our analytical model, which is based on the dilute solution approximation, along with the additional weak-field and small Dukhin assumptions, does not properly describe the well-known [28] experimentally observed decrease of the angular velocity magnitude with increasing electrolyte concentration (Fig.3b). Such a dependency, is still an open question in many experimental observations of ICEO and ACEO phenomena, where ionic crowding and charge-induced viscosity increase, were suggested as possible mechanisms [28]. Hence, we focus herein only on low conductivity (<1mM) electrolytes.

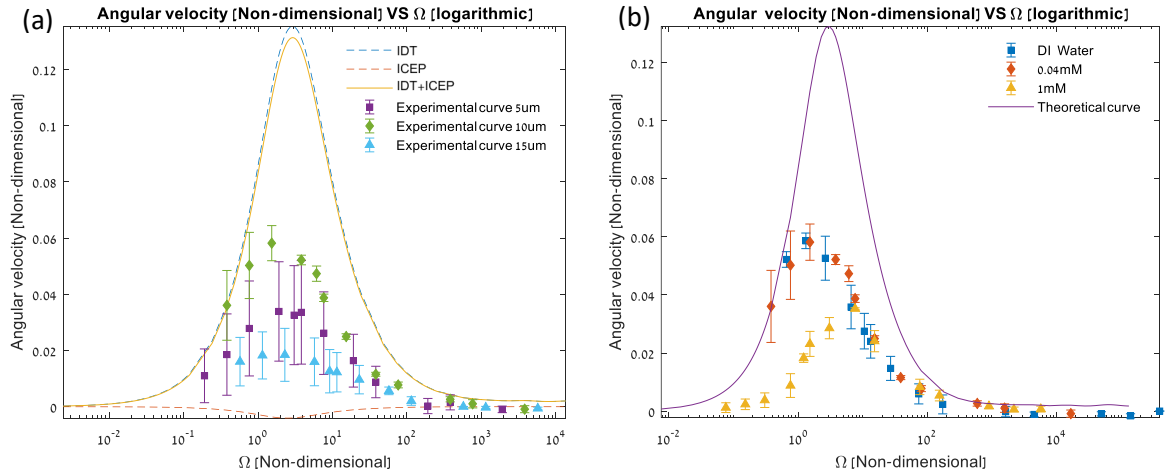

Equation (28) implies that the ICEP effect acts in a different direction (co-field) to the IDT contribution (counter-field) in the case of EROT. This might explain why the experimental results are smaller relative to the EROT theory. However, since it is plausible to use the same ICEP correction factor for both EOR and EROT, we have to choose a factor that will benefit either the EOR or the EROT. Since the ICEP effect is more distinguishable in EOR, we have determined this factor according to the EOR curves (Fig.4).

**Fig.3:** (a) **Analytical** (lines) and measured (symbols) angular velocities of 5,10, and 15 μm JPs immersed within 0.04mM KCl solution, under EROT conditions versus applied electric field frequency including ICEP, IDT and their combined ICEP+ IDT effect. Herein, a scaling factor of 0.1 was used for the ICEP contribution. (b) EROT spectra for DI water, 0.04mM and 1mM KCl solutions for 10 μm JPs.

### 4.3. Electro-orientation of JP

The EOR results are more intricate than the EROT, and each contribution (of the IDT and ICEP) has a separate frequency domain at which it is dominating. It can be deduced from the theory, that while the IDT contribution is significant at relatively high frequencies and is negligible at low frequencies, the ICEP contribution has an opposite behavior. When combined, these two contributions describe the full behavior of the JP EOR spectra. The ICEP contribution can explain the nonzero value observed at low frequencies. On the other hand, the IDT appears to dominate the high frequency response. In Fig.4. we show both the analytical and experimental Γ spectra (35), after normalizing it by $\varepsilon_e E_0^2 / \eta$, which again show a good qualitative agreement between the two. In case of homogenous spherical particles (e.g., semiconducting microspheres [3], gold coated sphere [9] etc.) there is no IDT related EOR torque since $\Re(d_\parallel - d_\perp) = 0$. However, for homogenous non-spherical particles (e.g., nanowires [11], and spheroids [12] $\Re(d_\parallel - d_\perp) \neq 0$, resulting in a clear EOR reaction. To our best knowledge there are no previous experimental investigations of EOR of JP. The EROT experimental data [8], [14], were obtained by us (compare to [8,14]), in order to complement our EOR results using exactly the same experimental setup and test conditions.

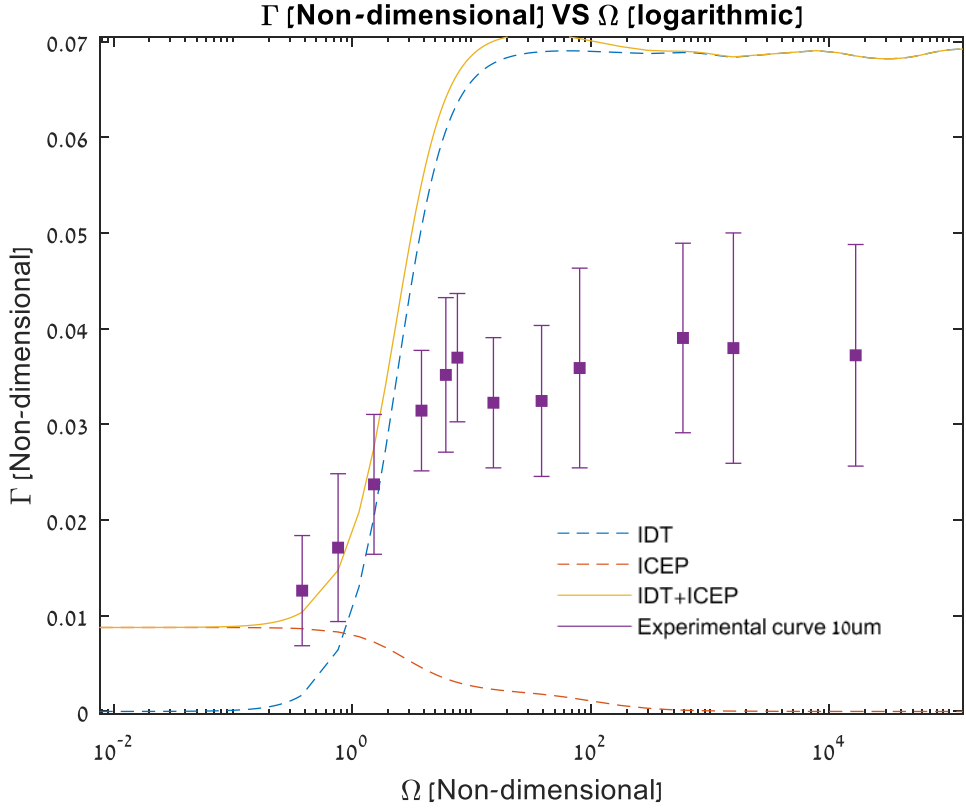

**Fig.4:** Calculated (lines) and measured (symbols) Γ (37) of the JP under electro-orientation (EOR) conditions versus applied electric field frequency including ICEP, IDT and their combined ICEP+ IDT effect. Herein, a scaling factor of 0.1 was used for the ICEP contribution.

## 5. Conclusions

This work presents a comprehensive theoretical and experimental study of the AC electrokinetic dynamic response of a spherical MD JP under the combined effect of EROT and EOR electric forcing, based on the 'weak field' assumption and a thin EDL approximation. To our best knowledge, this is the first obtained rigorous analytical solution for explicitly determining (to any degree of accuracy) the frequency-dependent (spectra) dipoles terms (including higher-order multipoles), resulting from material symmetry-breaking effects of the metallodielectrtic (MD) spherical JP under an inclined (relative to the metallodielectric interface) AC ambient electric field. Both normal and transverse dipoles exhibit a strong frequency dependence (dispersion), with the induced dipole in the parallel (to the interface) direction being much larger than the induced dipole in the orthogonal (perpendicular) direction. However, one cannot ignore the effect of the latter and its interaction with the former, especially when it comes to calculating EOR response. The angular rotation of a JP incited by the ambient AC electric fields, results from both electrostatic induced-dipole torque (IDT) and induced-charge electro-phoretic (ICEP) torques exerted on the particle.

In EROT, the out-of-phase AC fields generate a rotational JP motion governed by the imaginary part of the corresponding induced-dipoles. The asynchronous EROT motion is the result of the combined effects of both IDT and ICEP. While the IDT generally exerts a counter-field torque, the ICEP torque is acting in the opposite direction (co-field). Both IDT and ICEP torque spectra exhibit a Lorentzian behavior and monotonically vanish for $\omega \ll \omega_{RC}$, $\omega \gg \omega_{RC}$. This conclusion follows from the fact that the explicit expressions found for the ICEP and IDT torques, depend on the imaginary part of the induced dipoles and are verified here both analytically and experimentally, in accordance with previous studies [11], [14], [15], [21], [23].

As far as EOR dynamics, the in-phase AC fields generate a rotational motion governed by the real part of the induced dipoles. This motion is due to the combined effects of both IDT and ICEP. While the IDT effect seems to be dominating at the high-frequency range $\omega \gg \omega_{RC}$, the ICEP torque appears to dominate at the low-frequency range $\omega \ll \omega_{RC}$. By combining the two, one can thus predict the overall transient EOR rotation of the JP at any given frequency. It is also important to note that electroosmotic (ICEO) effects may contribute to overall EROT and EOR especially in the low-frequency domain. However, for thin EDL and uncharged particles (small zeta potential), the ICEO effects is negligibly small compared to IDT [24] [34], as well as the effect of natural surface charge on the polarizability of the JP's dielectric hemisphere that is negligible for the tested conditions [36] (see Fig.2 in ref. [36], wherein for zeta potential of ~-20mV, typical to polystyrene particles [37] from which the JP are made of, and for thin EDL ~50nm corresponding to 0.04mM KCl solution, the effect of the zeta potential on the particle polarizability if negligible).

Finally, it is important to note that the present analysis for determining the JP mobility and related spectra under different ambient uniform AC forcing, is based on using the first-order dipole approximation. Nevertheless, the accuracy of the newly proposed methodology can be systematically improved by also including higher-order multipoles. A possible interesting extension of the current analytical solution may be to account for more complicated (nonuniform) electric fields and the presence of nearby boundaries or interacting JPs (suspension), as well as considering the linear motion of a single free JP in addition to its rotation.

## 6. Materials and Methods
### 6.1. Device Fabrication

The quadrupolar electrode array substrate was fabricated using photolithography and wet-etching processes with distances across the center of electrodes of 500μm for EOR and EROT characterization. Layers of Au/Cr (200 nm/20 nm in thickness) were evaporated onto a glass substrate. Standard photolithography was then used to form the electrode pattern. The exposed Au/Cr surface was wet-etched. After the electrode array was fabricated, a flexible silicone chamber of 1mm in both diameter and depth (Grace BioLabs) was attached to the electrode substrate.

### 6.2. Janus particle and solution preparation

Fluorescently tagged polystyrene spheres (5, 10, and 15 μm in diameter) (Fluoro-max) in isopropanol (IPA) were pipetted onto a glass slide to form a monolayer upon solvent evaporation. The glass slide with particles was then coated with 10 nm Cr followed by 50 nm Nickel and then 30 nm (for the 10μm JPs) and 35 nm (for the 5μm and 15μm JPs) Au, following the protocol outlined in [35]. After then, JPs were rinsed three times with deionized water with 0.05 % non-ionic surfactant (Tween 20 (Sigma Aldrich)) in order to minimize non-specific adhesion onto the glass substrate and replaced with the target KCl solution. The solutions were made by diluting a concentrated KCl solution within DI water (obtained by Barnstead -Easypure II by Thermo Scientific) and adding 0.05% vol./vol. Tween 20. The conductivity was verified using a conductivity meter (B-173; Twin Cond by Horiba).

**Supporting Information**

Supporting information includes Figure S.1 depicting the convergence of the dipole calculation.

**Acknowledgments**

The fabrication of the chip was possible through the financial and technical support of the Technion RBNI (Russell Berrie Nanotechnology Institute) and MNFU (Micro Nano Fabrication Unit). This research was supported by BSF grant number 2018168.

**References**

[1] C. W. Shields, K. Han, F. Ma, T. Miloh, G. Yossifon, and O. D. Velev, "Supercolloidal Spinners: Complex Active Particles for Electrically Powered and Switchable Rotation," *Adv Funct Mater*, vol. 28, no. 35, Aug. 2018, doi: 10.1002/adfm.201803465.

[2] D. K. Sahu and S. Dhara, "Measuring electric-field-induced dipole moments of metal-dielectric janus particles in a nematic liquid crystal," *Phys Rev Appl*, vol. 14, no. 3, Sep. 2020, doi: 10.1103/PhysRevApplied.14.034004.

[3] L. Rodríguez-Sánchez, A. Ramos, and P. García-Sánchez, "Electrorotation of semiconducting microspheres," *Phys Rev E*, vol. 100, no. 4, Oct. 2019, doi: 10.1103/PhysRevE.100.042616.

[4] G. Schwarz, M. Saito, and H. P. Schwan, "On the orientation of nonspherical particles in an alternating electrical field," *J Chem Phys*, vol. 43, no. 10, pp. 3562–3569, 1965, doi: 10.1063/1.1696519.

[5] A. Boymelgreen, G. Yossifon, and T. Miloh, "Propulsion of Active Colloids by Self-Induced Field Gradients," *Langmuir*, vol. 32, no. 37, pp. 9540–9547, Sep. 2016, doi: 10.1021/acs.langmuir.6b01758.

[6] H. Morgan and N. G Green, "AC Electrokinetics: colloids and nanoparticles," Research Studies Press, 2003.

[7] Ronald. Pethig, "Dielectrophoresis : theory, methodology, and biological applications," Wiley, 2017.

[8] B. Behdani, K. Wang, and C. A. Silvera Batista, "Electric polarizability of metallodielectric Janus particles in electrolyte solutions," *Soft Matter*, 2021, doi: 10.1039/d1sm01046h.

[9] P. García-Sánchez, Y. Ren, J. J. Arcenegui, H. Morgan, and A. Ramos, "Alternating current electrokinetic properties of gold-coated microspheres," *Langmuir*, vol. 28, no. 39, pp. 13861–13870, Oct. 2012, doi: 10.1021/la302402v.

[10] C. H. Lin, Y. L. Chen, and H. R. Jiang, "Orientation-dependent induced-charge electrophoresis of magnetic metal-coated Janus particles with different coating thicknesses," *RSC Adv*, vol. 7, no. 73, pp. 46118–46123, 2017, doi: 10.1039/c7ra08527c.

[11] J. J. Arcenegui, P. García-Sánchez, H. Morgan, and A. Ramos, "Electro-orientation and electrorotation of metal nanowires," *Phys Rev E Stat Nonlin Soft Matter Phys*, vol. 88, no. 6, Dec. 2013, doi: 10.1103/PhysRevE.88.063018.

[12] T. Miloh and B. W. Goldstein, "Electro-phoretic rotation and orientation of polarizable spheroidal particles in AC fields," *Physics of Fluids*, vol. 27, no. 2, Feb. 2015, doi: 10.1063/1.4908527.

[13] P. Arenas-Guerrero, Á. v. Delgado, A. Ramos, and M. L. Jiménez, "Electro-Orientation of Silver Nanowires in Alternating Fields," *Langmuir*, vol. 35, no. 3, pp. 687–694, Jan. 2019, doi: 10.1021/acs.langmuir.8b03122.

[14] Y. L. Chen and H. R. Jiang, "Electrorotation of a metallic coated Janus particle under AC electric fields," *Appl Phys Lett*, vol. 109, no. 19, Nov. 2016, doi: 10.1063/1.4967740.

[15] J. Yan, M. Han, J. Zhang, C. Xu, E. Luijten, and S. Granick, "Reconfiguring active particles by electrostatic imbalance," *Nat Mater*, vol. 15, no. 10, pp. 1095–1099, Oct. 2016, doi: 10.1038/nmat4696.

[16] R. D. Miller and T. B. Jones, "Electro-orientation of ellipsoidal erythrocytes. Theory and experiment," *Biophys J*, vol. 64, no. 5, pp. 1588–1595, 1993, doi: 10.1016/S0006-3495(93)81529-7.

[17] S. Hwang *et al.*, "Cooperative Switching in Large-Area Assemblies of Magnetic Janus Particles," *Adv Funct Mater*, vol. 30, no. 26, Jun. 2020, doi: 10.1002/adfm.201907865.

[18] "I. N. Shklyarevskii, P. L. Pakhmov, USSR, Optika i Spektroskopiya 34 (1973)163, MIT database site.," 1973. http://www.mit.edu/~6.777/matprops/gold.htm (accessed Apr. 29, 2022).

[19] S. W. Ellingson, "Electromagnetics," vol. 1, Blacksburg, VA: VT Publishing. https://doi.org/10.21061/electromagnetics-vol-1 CC BY-SA 4.0, 2018.

[20] "MIT - material database site." http://www.mit.edu/~6.777/matprops/gold.htm (accessed Apr. 29, 2022).

[21] T. Miloh, "A unified theory of dipolophoresis for nanoparticles," in *Physics of Fluids*, 2008, vol. 20, no. 10. doi: 10.1063/1.2997344.

[22] A. M. Boymelgreen and T. Miloh, "Induced-charge electrophoresis of uncharged dielectric spherical Janus particles," *Electrophoresis*, vol. 33, no. 5, pp. 870–879, Mar. 2012, doi: 10.1002/elps.201100446.

[23] P. García-Sánchez and A. Ramos, "Electrorotation and Electroorientation of Semiconductor Nanowires," *Langmuir*, vol. 33, no. 34, pp. 8553–8561, Aug. 2017, doi: 10.1021/acs.langmuir.7b01916.

[24] P. García-Sánchez and A. Ramos, "Electrorotation of a metal sphere immersed in an electrolyte of finite Debye length," *Phys Rev E Stat Nonlin Soft Matter Phys*, vol. 92, no. 5, Nov. 2015, doi: 10.1103/PhysRevE.92.052313.

[25] M. Teubner, "The motion of charged colloidal particles in electric fields," *J Chem Phys*, vol. 76, no. 11, pp. 5564–5573, 1981, doi: 10.1063/1.442861.

[26] T. Jones, "Electromechanics of Particles.," in *Electromechanics of Particles*, Cambridge University Press, 1995, pp. 5–33. doi: 10.1017/CBO9780511574498.

[27] Q. Liu and A. Prosperetti, "Wall effects on a rotating sphere," *J Fluid Mech*, vol. 657, pp. 1–21, Aug. 2010, doi: 10.1017/S002211201000128X.

[28] M. Z. Bazant, M. S. Kilic, B. D. Storey, and A. Ajdari, "Towards an understanding of induced-charge electrokinetics at large applied voltages in concentrated solutions," *Advances in Colloid and Interface Science*, vol. 152, no. 1–2. pp. 48–88, Nov. 30, 2009. doi: 10.1016/j.cis.2009.10.001.

[29] J. J. Arcenegui, A. Ramos, P. García-Sánchez, and H. Morgan, "Electrorotation of titanium microspheres," *Electrophoresis*, vol. 34, no. 7, pp. 979–986, Apr. 2013, doi: 10.1002/elps.201200403.

[30] T. M. Squires and M. Z. Bazant, "Breaking symmetries in induced-charge electro-osmosis and electrophoresis," *J Fluid Mech*, vol. 560, pp. 65–101, 2006, doi: 10.1017/S0022112006000371.

[31] S. Gangwal, O. J. Cayre, M. Z. Bazant, and O. D. Velev, "Induced-charge electrophoresis of metallodielectric particles," *Phys Rev Lett*, vol. 100, no. 5, Feb. 2008, doi: 10.1103/PhysRevLett.100.058302.

[32] A. Ramos, P. García-Sánchez, and H. Morgan, "AC electrokinetics of conducting microparticles: A review," *Current Opinion in Colloid and Interface Science*, vol. 24. Elsevier Ltd, pp. 79–90, Aug. 01, 2016. doi: 10.1016/j.cocis.2016.06.018.

[33] H. Maier, "Electrorotation of colloidal particles and cells depends on surface charge," *Biophys J*, vol. 73, no. 3, pp. 1617–1626, 1997, doi: 10.1016/S0006-3495(97)78193-1.

[34] C. Grosse and V. N. Shilov, "Theory of the Low-Frequency Electrorotation of Polystyrene Particles in Electrolyte Solution," *J. Phys. Chem*., vol. 100 (5), pp. 1771–1778, 1996.

[35] A. Boymelgreen and G. Yossifon, "Observing Electrokinetic Janus Particle-Channel Wall Interaction Using Microparticle Image Velocimetry," *Langmuir*, vol. 31, no. 30, pp. 8243–8250, Aug. 2015, doi: 10.1021/acs.langmuir.5b01199.

[36] H. Zhou, M. A. Preston, R. D. Tilton, L. R. White, "Calculation of the electric polarizability of a charged spherical dielectric particle by the theory of colloidal electrokinetics," *Journal of Colloid and Interface Science*, vol. 285, pp. 845–856, 2005.

[37] B. J. Kirby, E. F. Hasselbrink, "Zeta potential of microfluidic substrates: 2. Data for polymers," *Electrophoresis*, vol. 25, pp. 203–213, 2004.

**For Table of Contents Only**

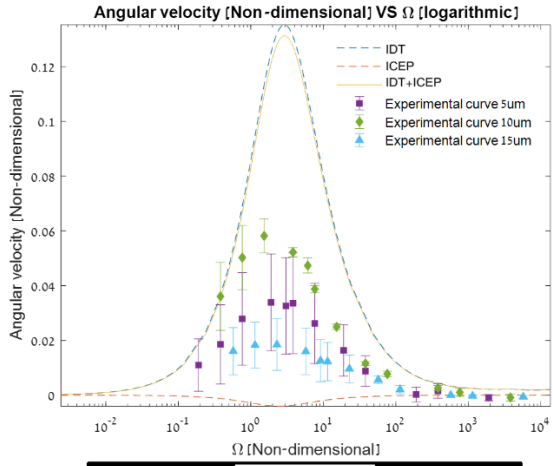
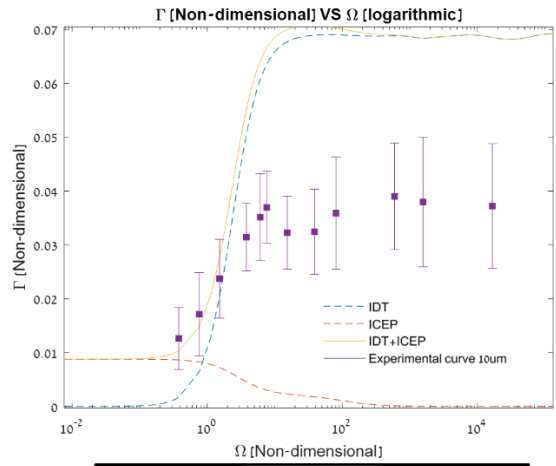
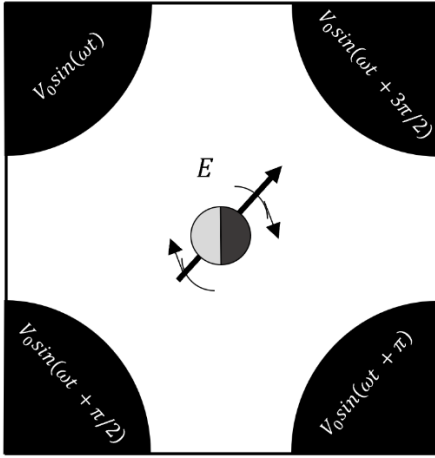
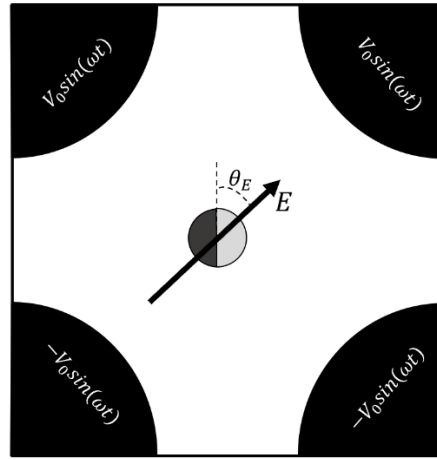